\documentclass[prc,superscriptaddress,unsortedaddress,twocolumn,showpacs,preprintnumbers,amsmath,amssymb]{revtex4}
\usepackage[dvips]{graphicx}
\usepackage{dcolumn}
\usepackage{amsmath}
\usepackage{times}

\def\fun#1#2{\lower3.6pt\vbox{\baselineskip0pt\lineskip.9pt
\ialign{$\mathsurround=0pt#1\hfil##\hfil$\crcr#2\crcr\sim\crcr}}}
\begin{document}

\title{Coulomb breakup effects on the elastic cross section\\ of
$^6$He+$^{209}$Bi scattering near Coulomb barrier energies}

\author{T. Matsumoto}
\email[Electronic address: ]{taku2scp@rarfaxp.riken.jp}
\affiliation{Institute of Physical and Chemical Research (RIKEN),
Hirosawa 2-1, Wako, Saitama 351-0198, Japan}

\author{T. Egami}
\affiliation{Department of Physics, Kyushu University, Fukuoka 812-8581,
Japan}

\author{K. Ogata}
\affiliation{Department of Physics, Kyushu University, Fukuoka 812-8581,
Japan}

\author{Y. Iseri}
\affiliation{Department of Physics, Chiba-Keizai College, Inage, Chiba
263-0021, Japan}

\author{M. Kamimura}
\affiliation{Department of Physics, Kyushu University, Fukuoka 812-8581,
Japan}

\author{M. Yahiro}
\affiliation{Department of Physics, Kyushu University, Fukuoka 812-8581,
Japan}

\date{\today}

\begin{abstract}
 We accurately analyze the $^6$He+$^{209}$Bi scattering at 19 and 22.5 MeV
 near the Coulomb barrier energy, using the continuum-discretized
 coupled-channels method (CDCC) based on the $n$+$n$+$^4$He+$^{209}$Bi
 four-body model.
 The three-body breakup continuum of $^6$He is discretized by diagonalizing
 the internal Hamiltonian of $^6$He in a space spanned by the Gaussian
 basis functions.
 The calculated elastic and total reaction
 cross sections are in good agreement with the experimental data,
 while the CDCC calculation based on the di-neutron model of $^6$He, i.e.,
 the $^2n$+$^{4}$He+$^{209}$Bi three-body model, does not reproduce the
 data.
\end{abstract}

\pacs{21.45.+v, 24.10.Eq, 25.60.-t, 25.70.De}

\maketitle

In the recent measurements of $^6$He+$^{209}$Bi scattering at 19 and 22.5 MeV
near the
Coulomb barrier energy~\cite{exp6He1,exp6He}, large enhancement of
the $\alpha$-emission cross section,
which dominated the total reaction cross section,
compared with that for the corresponding $^6$Li-induced reactions was
reported.
In order to clarify the nature of the enhancement,
Keeley \textit{et al.}~\cite{Keeley} analyzed the scattering
by means of
the continuum-discretized coupled-channels method 
(CDCC)~\cite{cdcc,cdcc1,sakuragi}
that is a fully quantum-mechanical method for describing scattering
of a three-body system.
In the analysis, the $^6$He+$^{209}$Bi system was
assumed to be the $^2n$+$^{4}$He+$^{209}$Bi three-body system,
that is, the neutron pair in $^6$He was treated as
a single particle, di-neutron ($^2n$). They found that the enhancement of
the total reaction cross section of the $^6$He+$^{209}$Bi scattering was
due to the electric dipole ($E1$) excitation of $^6$He to its continuum
states, i.e. Coulomb breakup processes of $^6$He, which
was approximately absent in the $^6$Li+$^{209}$Bi scattering case.
Their calculation, however,
did not reproduce the angular distribution of the measured
elastic cross section and overestimated the measured
total reaction cross section by a factor of three.
Thus, reaction mechanisms of the $^6$He+$^{209}$Bi scattering are not fully
understood.

In the very recent work~\cite{Rusek}, it was reported that the
elastic cross sections of the $^6$He+$^{209}$Bi scattering
calculated within the same framework as in Ref.~\cite{Keeley}
with the strength of the dipole coupling potentials multiplied by 0.5
reproduced the experimental data.
This indicates that the $E1$ excitation strength of $^6$He cannot be
accurately reproduced by the $^2n$+$^4$He model.
Since $^6$He is well known as a two-neutron halo nucleus, its structure
should be described by the $n$+$n$+$^4$He three-body system rather than
the $^2n$+$^{4}$He two-body one.
Thus, it is necessary to analyze the $^6$He+$^{209}$Bi scattering
by using the $n$+$n$+$^4$He+$^{209}$Bi four-body model.
Furthermore, a fully quantum-mechanical method such as CDCC is
highly required to analyze the scattering near the Coulomb
barrier energies in which
both nuclear and Coulomb breakup processes are significant.

In our previous work~\cite{4B-CDCC}, we proposed four-body CDCC
that is the extension of CDCC and describes four-body breakup processes.
In four-body CDCC, the three-body breakup continuum of the projectile
is discretized by diagonalizing the internal Hamiltonian 
in a space spanned by the Gaussian basis functions.
So far the Gaussian basis function was
used with success for solving bound-state problems of few-body systems.
The approach is called the Gaussian expansion method (GEM)~\cite{GEM}.
The application of the Gaussian basis function to the discretization of
breakup continuum is a natural extension of GEM. In general, the method
that describes the breakup continuum by a superposition of $L^2$-type
basis functions is called the pseudostate method~\cite{pscdc}, and other
basis functions have been also proposed so far~\cite{PS,THO1,THO2}.
Four-body CDCC was successfully applied to $^6$He+$^{12}$C
scattering at 18 and 229.8 MeV in which only nuclear breakup
was significant.
The elastic and breakup cross sections calculated with four-body CDCC
are found to converge as the number of Gaussian basis functions is
increased~\cite{4B-CDCC}.
This indicates that the set of discretized continuum states obtained
by the pseudostate method with GEM forms an complete set with good accuracy
in a finite region of space that is important for the four-body reaction
process concerned.
Furthermore, in Ref.~\cite{Egami}, applicability of CDCC
with the pseudostate method to Coulomb breakup processes, in which
large modelspace of the projectile is required because of the
long-range property of Coulomb coupling potentials, was shown
for the $^8$B+$^{58}$Ni scattering at 25.8 MeV.
Thus, it is expected that four-body CDCC with GEM accurately describes
the $^6$He+$^{209}$Bi scattering near the Coulomb barrier energy.

In this Rapid Communication, we analyze $^6$He+$^{209}$Bi scattering at
19 and 22.5 MeV by means of
four-body CDCC. This is the first application of
four-body CDCC to low-energy scattering in which both nuclear and Coulomb
breakup processes are significant. We show that four-body CDCC
reproduces the measured elastic and total reaction cross sections reasonably
well. Effects of four-body breakup processes on the elastic scattering are
investigated through the dynamical polarization (DP) potential.
We discuss the reason why the di-neutron model of $^6$He is insufficient
to describe the $^6$He+$^{209}$Bi scattering.

\begin{figure}[htbp]
\begin{center}
 \includegraphics[width=0.36\textwidth,clip]{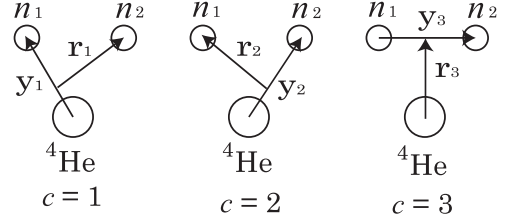}
 \caption{Jacobi coordinates of the three rearrangement channels
 ($c=1\mbox{--}3$) adopted for the $n$+$n$+$^4$He model of $^6$He.
 }
 \label{Jacobi}
\end{center}
\end{figure}

We assume that the $^6$He+$^{209}$Bi scattering is described as
the $n$+$n$+$^4$He+$^{209}$Bi four-body system.
The model Hamiltonian of the system is defined by
\begin{eqnarray}
 H=K_{R}+U_{n{\rm Bi}}(R_{n_1})+U_{n{\rm Bi}}(R_{n_2})
  +U_{\alpha{\rm Bi}}(R_{\alpha})
  +H_6,
\label{H4}
\end{eqnarray}
where ${\bf R}_{X}$ (${\bf R}$) is the coordinate of
particle $X$ (the center-of-mass of $^{6}$He) relative
to $^{209}$Bi,
$K_{R}$ is the kinetic energy associated with ${\bf R}$, and $H_6$
is the internal Hamiltonian of $^6$He. The potential $U_{n{\rm Bi}}$
($U_{\alpha{\rm Bi}}$) represents the interaction between $n$
($^4$He) and $^{209}$Bi. It should be noted that
$U_{\alpha{\rm Bi}}$ contains a Coulomb part that causes
Coulomb breakup processes in the $^6$He+$^{209}$Bi scattering.
In four-body CDCC, the total wave function of the four-body system is
expanded in terms of a finite number of the internal wave functions of
the $^6$He projectile. The internal wave functions including the bound
and discretized-continuum states are generated with GEM as mentioned above.
In GEM, the $n$ th eigenstate $\Phi_{nIm}$ of
$^6$He with the total spin $I$ and its projection on the $z$-axis $m$ is
written as
\begin{equation}
\Phi_{nIm}={\sum_{c=1}^3} \psi^{(c)}_{nIm}({\bf y}_c,{\bf r}_c),
\label{phig}
\end{equation}
where $c$ denotes a set of Jacobi coordinates shown in
Fig.~\ref{Jacobi}. Each $\psi^{(c)}_{nIm}$ is then expanded in terms
of the Gaussian basis functions:
\begin{eqnarray}
&&\hskip -0.3cm \psi^{(c)}_{nIm}({\bf y}_c,{\bf r}_c)=
\sum_{\lambda,\ell,\Lambda,S}
\sum_{i=1}^{i_{\rm max}}
\sum_{j=1}^{j_{\rm max}}
  A_{i\lambda j\ell\Lambda S}^{(c)nI}\nonumber \\
&& \hskip -0.3cm \times
  y_c^{\lambda}\,r_c^{\ell}\,
  e^{-({y_c}/{\bar{y}_{i}})^2}\,
  e^{-({r_c}/{\bar{r}_{j}})^2} \nonumber \\
&& \hskip -0.3cm \times
   \left[
   \left[
    Y_{\lambda}(\hat{\bf y}_c)
    \otimes Y_{\ell}(\hat{\bf r}_c)
  \right]_\Lambda
\otimes
\left[
\eta_{\frac{1}{2}}^{(n_1)}\otimes\eta_{\frac{1}{2}}^{(n_2)}
\right]_{S}
\right]_{Im},
\label{gauex}
\end{eqnarray}
where $\lambda$ ($\ell$) is the angular momentum
regarding the Jacobi coordinate ${\bf y}_c$ (${\bf r}_c$),
and $\eta_{1/2}$ is the spin wave function of each valence neutron
($n_1$ or $n_2$). In actual calculation we truncate $\lambda$ and
$\ell$ at appropriate maximum values, $\lambda_{\rm max}$ and
$\ell_{\rm max}$, respectively.
The Gaussian range parameters are taken to lie in
geometric progression:
\begin{eqnarray}
 \bar{y}_{i}=\bar{y}_1
  (\bar{y}_{\rm max}/\bar{y}_1)^{(i-1)/(i_{\rm max}-1)},
\label{range1}
 \\
 \bar{r}_{j}=\bar{r}_1 (\bar{r}_{\rm max}/
  \bar{r}_1)^{(j-1)/(j_{\rm max}-1)}.
\label{range2}
\end{eqnarray}
The eigenstate $\Phi_{nIm}$ of $^6$He is
antisymmetrized for the exchange between $n_1$ and $n_2$; we then have
$A_{i\lambda j\ell\Lambda S}^{({2})nI}=(-)^SA_{i\lambda j\ell\Lambda S}^{({1})nI}$ and $(-)^{\lambda+S}=1$
for $c=3$. Meanwhile, the exchange between each valence neutron and each
nucleon in $^4$He is treated approximately by the orthogonality
condition model~\cite{OCM}. The eigenenergies $\epsilon_{nI}$ of $^6$He
and the corresponding expansion-coefficients
$A_{i\lambda j\ell\Lambda S}^{(c)nI}$ are
determined by diagonalizing $H_6$~\cite{GEM6He1,GEM6He}.

Using the internal states of $^6$He thus obtained, we expand the total
wave function of the $n$+$n$+$^4$He+$^{209}$Bi four-body system with
the total angular momentum $J$ and its projection on the $z$-axis $M$,
$\Psi^{JM}$:
\begin{eqnarray}
 \Psi^{JM}=\sum_{nIL}
\chi_{nIL}^J (P_{nI},R)/R
\;{\cal Y}_{nIL}^{JM},
\end{eqnarray}
${\cal Y}_{\gamma}^{JM}=[\Phi_{nI}({\bf y}_c,
{\bf r}_c)\otimes i^LY_L(\hat{\bf R})]_{JM}$ and
$L$ is the orbital angular momentum regarding ${\bf R}$;
below we denotes the channels \{$n,I,L$\} as $\gamma$.
The expansion-coefficient $\chi_{\gamma}^J$
represents the relative motion between the $^6$He projectile and the
$^{209}$Bi target and $P_{nI}$ is the corresponding relative momentum.
Multiplying the four-body
Schr\"{o}dinger equation $(H-E)\Psi^{JM}=0$ by
${\cal Y}_{\gamma'}^{*JM}$ from the left and integrating over
all variables except $R$, one obtains the set of coupled
differential equations for $\chi_{\gamma}^J$
\begin{eqnarray}
\bigg[\frac{d^2}{dR^2}
-\frac{L(L+1)}{R^2}
-\frac{2\mu}{\hbar^2}U_{\gamma\gamma}(R)
+P_{nI}^2
\bigg]
\chi_{\gamma}^J(P_{nI},R)
\nonumber \\
&&\hspace{-55mm}
=
\frac{2\mu}{\hbar^2}\sum_{\gamma'\ne \gamma}
U_{\gamma'\gamma}(R)\chi_{\gamma'}^J(P_{n'I'},R),
\label{CDCCeq}
\end{eqnarray}
where the coupling potential $U_{\gamma'\gamma}(R)$ is defined by
\begin{equation}
U_{\gamma'\gamma}(R)=
\langle {\cal Y}_{\gamma'}^{JM}|
U_{n{\rm Bi}}(R_{n_1})+U_{n{\rm Bi}}(R_{n_2})
+U_{\alpha{\rm Bi}}(R_{\alpha})
|{\cal Y}_{\gamma}^{JM} \rangle
\nonumber
\end{equation}
and $\mu$ is the reduced mass between $^6$He and $^{209}$Bi.
We obtain the elastic and discrete breakup $S$-matrix elements
by solving Eq.~(\ref{CDCCeq}) under an appropriate asymptotic
boundary condition~\cite{cdcc,Piya}.
Details of the formalism of CDCC are shown in Ref.~\cite{cdcc}.

\begin{table}[htbp]
 \caption{The maximum internal angular momenta and the Gaussian
 range parameters for each Jacobi coordinate.}
 \begin{tabular}{cccccccc}
  \hline
  $c$&$I^\pi$&$\lambda_{\rm max}$&$\ell_{\rm max}$&
  $\bar{y}_1$ [fm]&$\bar{y}_{\rm max}$ [fm]
  &$\bar{r}_1$ [fm]&$\bar{r}_{\rm max}$ [fm]
  \\ \hline\hline
  3&$0^+$ & 1&1&  0.1&       15.0&    0.5&  15.0   \\
  1,2&$0^+$ & 1&1&  0.5&       15.0&    0.5&  15.0 \\ \hline
  3&$1^-$ & 1&1&  0.1&       15.0&    0.5&  15.0   \\
  1,2&$1^-$ & 1&1&  0.5&       15.0&    0.5&  15.0 \\ \hline
  3&$2^+$ & 2&2&  0.1&       15.0&    0.5&  15.0   \\
  1,2&$2^+$ & 1&1&  0.5&       15.0&    0.5&  15.0 \\ \hline
 \end{tabular}
 \label{tab1}
\end{table}

In the present four-body CDCC calculation for $^6$He+$^{209}$Bi
scattering at 19 and 22.5 MeV, we take $I^\pi=0^+$, $1^-$, and $2^+$
states for $^6$He. Inclusion of $1^-$ state is essential to
describe Coulomb breakup processes. As for the internal Hamiltonian of $^6$He,
we adopt the same Hamiltonian as used in Ref.~\cite{4B-CDCC}.
We show in Table~\ref{tab1} the maximum values of the internal angular
momenta, $\lambda_{\rm max}$ and $\ell_{\rm max}$,
and the Gaussian range parameters, $\bar{y}_1$, $\bar{y}_{\rm max}$,
$\bar{r}_1$, and $\bar{r}_{\rm max}$, used in the calculation of
$\Phi_{nIm}$.
For each set of $\{c,\lambda,\ell,\Lambda,S\}$,
we take $i_{\rm max}=j_{\rm max}=10$.
These values of the parameters are found to give good convergence
of the calculated elastic and total reaction cross sections.
It should be noted that
the maximum value of each Gaussian range parameter is 15 fm,
which is quite larger than that used in the four-body CDCC
analysis of $^6$He nuclear breakup~\cite{4B-CDCC}, i.e. 10 fm.
Some parameters shown in Table \ref{tab1} depend on $I^\pi$ and $c$,
while in
Eqs.~(\ref{gauex})--(\ref{range2}) the dependence has not been shown for
simplicity.

We select the $\Phi_{nIm}$ with $\epsilon_{nI} \le 7$ MeV among those
obtained by diagonalizing $H_6$ and use them in actual CDCC calculation,
since the $\Phi_{nIm}$ with $\epsilon_{nI} > 7$ MeV are found to
have no effect on the calculated cross sections of the $^6$He+$^{209}$Bi
scattering shown below.
The resulting number of the discrete states
for the $0^+$, $1^-$, and $2^+$ states is
37 (including the ground state of $^6$He),
44, and 53, respectively.
As for the nuclear parts of $U_{n{\rm Bi}}$ and $U_{\alpha{\rm Bi}}$,
respectively, we take
the optical potentials of Koning and Delaroche~\cite{n-Bi} and of Barnett
and Lilley~\cite{a-Bi}.
The maximum value of $L$ is taken to be 200 and the scattering wave
function $\chi_{\gamma}^J$ is connected to its appropriate asymptotic
form at $R=200$ fm.

Below we show also results of CDCC calculation
based on the $^2n$+$^{4}$He+$^{209}$Bi three-body model
to see three-body breakup effects on the elastic and
total reaction cross sections.
In this model the di-neutron ($^2n$) model of $^6$He is used
and the intrinsic spin and the relative energy of $^2n$ are assumed
to be zero. We henceforth call CDCC based on the model above
three-body CDCC.
As for the interaction between $^2n$ and $^4$He, we
take the same parameter as used in Ref.~\cite{Keeley}.
The $^2n$+$^4$He continuum is discretized by the pseudostate method
described in Ref.~\cite{pscdc}
and truncated at the relative momentum $k=0.7$ fm$^{-1}$ that
corresponds to about 7 MeV of the excitation energy of $^6$He.
The number of the discrete state is 9 for each of
the $0^+$, $1^-$, and $2^+$ states of $^6$He.
The range parameters of the Gaussian basis functions are
$(a_1=0.5~\mathrm{fm}, a_{n_a}=20.0~\mathrm{fm}, n_a=20 )$
with the same notation as in Ref.~\cite{pscdc}.
We adopt the optical potential of Barnett and Lilley~\cite{a-Bi} for
the nuclear part of $U_{\alpha{\rm Bi}}$, as in the four-body CDCC
calculation, and the $d$-$^{208}$Pb (type-a) optical potential
at the deuteron incident energy of 15.0 MeV~\cite{Perey}
for the interaction between $^2n$ and $^{209}$Bi.
Other parameters of the modelspace are the same as in the four-body
CDCC calculation.

\begin{figure}[htbp]
\includegraphics[width=0.35\textwidth,clip]{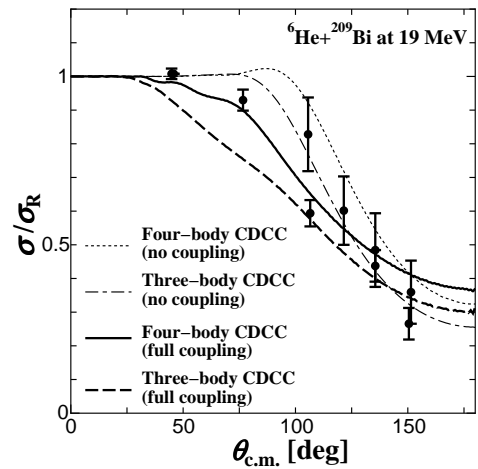}
\caption{Angular distribution of the elastic differential cross section
 as the ratio to the Rutherford cross section for
 the $^6$He+$^{209}$Bi scattering at
 19 MeV. The solid (dashed) and dotted (dot-dashed)
 lines show the results of the four-body CDCC (three-body CDCC)
 calculation with and without breakup effects, respectively.
 The experimental data are taken from Ref.~\cite{exp6He1,exp6He}.
 The incident energy for the experimental data in the laboratory frame
 is shown to be 19 MeV and 18.4 MeV in the first~\cite{exp6He1}
 and second~\cite{exp6He} papers of Aguilera \textit{et al.},
 respectively; in the present study we take 19 MeV.}
\label{elastic1}
\end{figure}

 \begin{figure}[htbp]
\includegraphics[width=0.35\textwidth,clip]{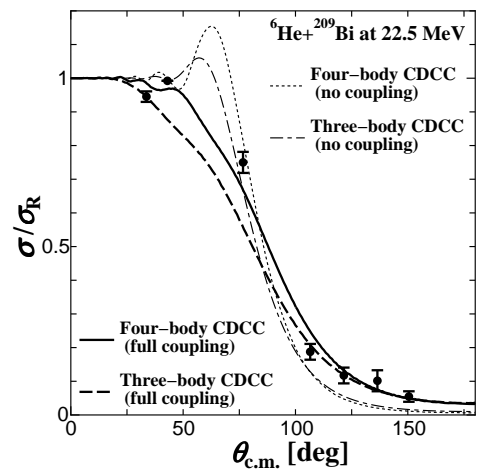}
\caption{The same as in Fig.~\ref{elastic1} but for $^6$He+$^{209}$Bi
 scattering at 22.5 MeV. The experimental data are taken from
 Ref.~\cite{exp6He1,exp6He}. We take the incident energy of 22.5 MeV
 shown in the first paper of Aguilera \textit{et al.}~\cite{exp6He1}.}
\label{elastic2}
 \end{figure}

Figure~\ref{elastic1} shows the angular distribution of the elastic
differential cross section for the $^6$He+$^{209}$Bi scattering at 19
MeV. The solid line is the result of four-body CDCC
and the dashed line is that of three-body CDCC.
The results of four-body CDCC and three-body CDCC
without breakup effects of $^6$He are shown by the
dotted and dot-dashed lines, respectively.
The difference between the solid and dotted (dashed and
dot-dashed) lines shows effects of the four-body (three-body)
breakup on the elastic cross section.
One sees that four-body CDCC reproduces
the experimental data well, while three-body CDCC
underestimates the data at middle angles of 50$^\circ$--100$^\circ$.
The dashed line is consistent with the result of
Ref.~\cite{Keeley}; it should be noted that the real part of each
coupling potential was multiplied by 0.8 in Ref.~\cite{Keeley}, while
in the present study such a renormalization factor is not included.
Figure~\ref{elastic2} shows the result at 22.5 MeV and features of the
result are just the same as at 19 MeV.

\begin{figure}[htbp]
\includegraphics[width=0.35\textwidth,clip]{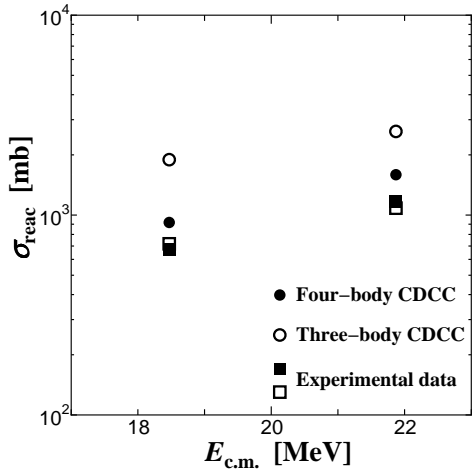}
\caption{Total reaction cross sections of $^6$He+$^{209}$Bi scattering
 as a function of the incident energy in the center of mass frame.
 The four values on the left (right) side correspond to the $^6$He
 incident energy of 19 MeV (22.5 MeV) in the laboratory frame.
 The solid and open circles represent the results of
 the four-body CDCC calculation and the three-body CDCC calculation,
 respectively. The experimental data shown by the open and
 solid squares are taken from Ref.~\cite{exp6He1}; see the text for details.}
\label{reac}
\end{figure}

We show in Fig.~\ref{reac} the calculated total reaction cross sections;
the solid (open) circles represent the results of four-body
(three-body) CDCC.
The open squares are the experimental values of the
$\alpha$-emission cross sections~\cite{exp6He1} added by
the fusion cross sections~\cite{Kolata}, while the solid squares
are the total reaction cross sections
evaluated from an optical-model analysis of the measured
elastic cross sections~\cite{exp6He1}.
The figure shows that three-body CDCC overestimates the experimental
data by a factor of about three.
This overestimation is consistent with the fact that
three-body CDCC underestimates the elastic cross section as shown in
Figs.~\ref{elastic1} and \ref{elastic2}.
On the contrary, four-body CDCC reproduces the experimental data
quite well, which shows the importance of the accurate description
of three-body breakup continuum of $^6$He.
The remaining difference of about a few tens of \% between
the results of four-body CDCC and the data needs further investigation,
including analysis of $^6$He+$^{208}$Pb scattering in barrier-energy
region~\cite{Pb-target}.

In order to clarify the reason why the total reaction cross
section calculated with three-body CDCC is much larger than that
with four-body CDCC, we see first the strength of E1 transition based
on the two models.
The non-energy weighted $E1$ excitation strength $B(E1)$
from the ground state $\Phi_{000}$ of $^6$He to its excited states
$\Phi_{nIm}$ with $I=1$ is given by
\begin{eqnarray}
 B(E1)=\left(\frac{2}{3}\right)^2
  \sum_{n}\sum_{\mu,m}
  \left|\langle \Phi_{n1m}|y_3Y_{1\mu}(\hat{\bf y}_3)|\Phi_{000}
   \rangle\right|^2.
\end{eqnarray}
The summation over $n$ is taken up to a value that corresponds to
the excitation energy of $^6$He of 7 MeV, i.e. the maximum energy
of the modelspace of the present analysis.
The resulting values of $B(E1)$ based on the
$^2n$+$^{4}$He model and the $n$+$n$+$^4$He model are, respectively,
1.5 $e^2{\rm fm}^2$ and 0.9 $e^2{\rm fm}^2$.
The latter agrees well with the experimental value reported by
Aumann \textit{et al.}~\cite{Aumann}, which is consistent with
the conclusion in Ref.~\cite{Thompson}.
Thus, the $E1$ strength is overestimated in the di-neutron model. This
indicates that the di-neutron model overshoots the breakup cross section
of the $^6$He+$^{209}$Bi scattering, which is the main reason why three-body
CDCC overestimates the result of four-body CDCC, hence the measured
total reaction cross section.
This conclusion is qualitatively consistent with that drawn in
Ref.~\cite{Rusek},
in which, as mentioned above, three-body CDCC with the strength of the dipole
coupling potentials multiplied by 0.5 was shown to reproduce the
elastic scattering data of the $^6$He+$^{209}$Bi scattering.

Next we discuss the difference of the optical potentials taken in
the $^2n$+$^4$He+$^{209}$Bi three-body model
and the $n$+$n$+$^4$He+$^{209}$Bi four-body model.
As one sees in Figs.~\ref{elastic1} and \ref{elastic2},
the elastic cross section without breakup effects calculated with the
three-body model underestimates that with the four-body model.
This means that the diagonal component of the imaginary potential
in the elastic channel based on the three-body model is deeper than
that on the four-body model.
Thus, the three-body model yields an absorption cross
section larger than the four-body model does.
This is also an important factor for the enhancement of the total
reaction cross section calculated with three-body CDCC.

Finally, we see the difference between three-body CDCC and
four-body CDCC in more detail by evaluating
the dynamical polarization (DP)
potential $U_{\rm DP}^J$ and the equivalent local potential
$U_{\rm eq}^{J}(R)$. Explicit form of the two potentials is given by
\begin{eqnarray}
 U_{\rm DP}^J(R)=
  \frac{
  \sum_{\gamma \ne \gamma_0}U_{\gamma\gamma_0}(R)\chi_{\gamma}^J(P_{nI},R)
  }
  {\chi_{\gamma_0}^{J}(P_{0I_0},R)}
 \label{Eq:DP}
\end{eqnarray}
and
\[
U_{\rm eq}^{J}(R)\equiv U_{\rm DP}^J(R)+U_{\gamma_0\gamma_0}(R),
\]
where the subscript 0 denotes the incident channel.

\begin{figure}[htbp]
\includegraphics[width=0.35\textwidth,clip]{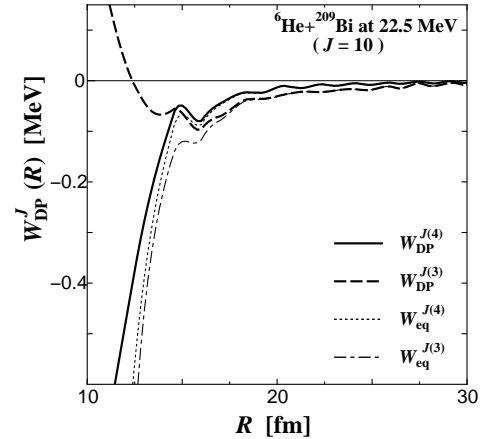}
\caption{The imaginary parts of the dynamical polarization (DP)
 potential and the equivalent local potential for the $^6$He+$^{209}$Bi
 scattering at 22.5 MeV with $J=10$.
 The solid and dotted lines, respectively, represent the results of
 the DP and equivalent local potential calculated with four-body CDCC.
 The dashed and dotted-dash lines correspond to those calculated with
 three-body CDCC based on the di-neutron model of $^6$He.}
\label{DPP}
\end{figure}

Figure~\ref{DPP} shows the imaginary parts $W_{\rm DP}^{J(4)}$ and
$W_{\rm DP}^{J(3)}$ of $U_{\rm DP}^J$ calculated with four-body CDCC
(solid line) and three-body CDCC (dashed line), respectively, for the
$^6$He+$^{209}$Bi scattering at 22.5 MeV. Also shown are the results of
the imaginary parts of $U_{\rm eq}^J$ calculated with four-body and
three-body CDCC, i.e. $W_{\rm eq}^{J(4)}$ and $W_{\rm eq}^{J(3)}$.
It should be noted that the figure shows the potentials only in the
peripheral region, i.e., 10 fm $\le R \le$ 30 fm, where the scattering
wave has nonnegligible values.
The results correspond to $J=10$ at which the partial reaction cross
section becomes maximum. It is confirmed that the DP potential hardly
depends on $J$ around $J=10$ in the peripheral region shown in
Fig.~\ref{DPP}. 
It should be noted that $U_{\rm DP}^J$ defined by Eq.~(\ref{Eq:DP}),
or $U_{\rm eq}^{J}(R)$, is
not a smooth function since it is divided by the oscillating function
$\chi_{\gamma_0}^J$.
Each line in Fig.~\ref{DPP} is obtained by interpolation
of $U_{\rm DP}^{J}(R)$ (or $U_{\rm eq}^{J}(R)$) at $R$ where
$|\chi_{\gamma_0}^J|$ is more than 90\% of its maximum value in the
asymptotic region; this simple way of evaluating $U_{\rm DP}^{J}(R)$ and
$U_{\rm eq}^{J}(R)$ is appropriate for the present purpose. The
$S$-matrix element calculated with the interpolated DP potential
reproduces the one obtained by CDCC within the relative error of 3\%.

One sees from Fig.~\ref{DPP}
that both $W_{\rm DP}^{J(3)}$ and $W_{\rm DP}^{J(4)}$ have a
long ranged tail, which is induced by Coulomb breakup processes.
In the tail region of $R\geq 15~{\rm fm}$,
where the imaginary part of $U_{\gamma_0\gamma_0}$ is negligible
and $W_{\rm DP}^{J}$ agrees with $W_{\rm eq}^{J}$,
$W_{\rm DP}^{J(3)}$ is deeper than $W_{\rm DP}^{J(4)}$.
This is consistent with the fact mentioned above
that the strength of E1 transition
is overestimated by the di-neutron model of $^6$He, which is used in
the three-body CDCC calculation.
On the other hand,
$W_{\rm DP}^{J(3)}$ is shallower than $W_{\rm DP}^{J(4)}$
in the region of $R \leq 15~{\rm fm}$.
However, the imaginary part of $U_{\gamma_0\gamma_0}$ calculated with
three-body CDCC is, as mentioned above,  much deeper than that with
four-body CDCC, which makes
$W_{\rm eq}^{J(3)}$ deeper than $W_{\rm eq}^{J(4)}$ in this region.
Thus, the imaginary part of the equivalent local potential,
which dictates the total reaction cross section, calculated with
three-body CDCC is deeper than that with four-body CDCC in the
entire region that is important for the $^6$He+$^{209}$Bi scattering
concerned.
This conclusion is consistent with the discussion above and
the result shown in Fig.~\ref{reac}.

The real part $V_{\rm DP}^{J}$
of the DP potential also plays important roles
in the $^6$He+$^{209}$Bi scattering. It is found that
$V_{\rm DP}^{J}$ is repulsive for $R\leq 15~{\rm fm}$,
where nuclear breakup processes are significant,
and attractive for $R\geq 15~{\rm fm}$, where Coulomb breakup processes
are dominant. The difference between $V_{\rm DP}^{J}$ calculated with
three-body CDCC and four-body CDCC in the outer region is just the
same as for $W_{\rm DP}^{J}$ shown in Fig.~\ref{DPP}.


In summary, the $^6$He+$^{209}$Bi scattering at 19 MeV and 22.5 MeV,
near the Coulomb barrier energy, is analyzed with
the continuum-discretized coupled-channels method (CDCC)
based on the $n$+$n$+$^4$He+$^{209}$Bi four-body model,
four-body CDCC, that treats both nuclear and Coulomb breakup processes
simultaneously.
It is found that the angular distribution of the elastic cross section
and the total reaction cross section calculated with four-body CDCC
reasonably reproduce the experimental data with no free adjustable
parameter.
Three-body CDCC based on the
$^2n$+$^4$He+$^{209}$Bi three-body model, i.e. di-neutron model of $^6$He,
turns out to overestimate the total reaction cross section by
a factor of about three, which is consistent with the conclusion of
the previous work of Keeley {\it et al.}~\cite{Keeley}.
The value of $B(E1)$ calculated by the
di-neutron model overestimates the experimental one, while
the $n$+$n$+$^4$He three-body model reproduces the data well,
as pointed out in Ref.~\cite{Rusek}.
This makes the imaginary part of the dynamical polarization
potential calculated with three-body CDCC deeper than that
with four-body CDCC in the tail region. Also important fact is
that three-body CDCC contains a deep imaginary part of the diagonal
potential for the elastic channel compared with four-body CDCC does.
We thus conclude that the di-neutron model is inadequate to treat
breakup continuum of $^6$He precisely and a $n$+$n$+$^4$He+$^{209}$Bi
four-body reaction model is necessary to accurately describe the
$^6$He+$^{209}$Bi scattering. Four-body CDCC is indispensable to analyze
low energy $^6$He scattering in which both nuclear and Coulomb breakup
processes are significant.

The authors would like to thank E. Hiyama for helpful discussions on the
GEM calculation. T. M. is grateful for financial assistance from the
Special Postdoctoral Researchers Program of RIKEN.
This work has been supported in part by the Grants-in-Aid for
Scientific Research of Monbukagakusyou of Japan and JSPS.


\end{document}